\documentclass[a4paper,11pt]{article}
\usepackage{jheppub} % for details on the use of the package, please see the JINST-author-manual
\usepackage{lineno}
\usepackage{cleveref}
\graphicspath{{figures/}}

\title{Chaos in the near-horizon dynamics of the dyonic \texorpdfstring{$\bf{AdS_4}$}{AdS4}-Reissner-Nordstr\"{o}m black hole}

\author[a]{Mu-Yang Wang,}
\author[b]{Si-Wen Li,}
\author[c]{Defu Hou,}
\author[a]{Dong Yan,}
\author[a]{Yan-Qing Zhao}

\affiliation[a]{College of Physics and Electronic Engineering, Hainan Normal University ,\\ Haikou 571158, China}
\affiliation[b]{Department of Physics, College of Science, Dalian Maritime University,\\ Dalian 116026, China}
\affiliation[c]{Institute of Particle Physics and Key Laboratory of Quark and Lepton Physics (MOS), \\Central China Normal University,\\ Wuhan 430079, China}

\emailAdd{wmy228@hainnu.edu.cn}
\emailAdd{siwenli@dlmu.edu.cn}
\emailAdd{houdf@mail.ccnu.edu.cn}
\emailAdd{yand@hainnu.edu.cn}
\emailAdd{zhaoyanqing@hainnu.edu.cn}

\abstract{We investigate the chaos in the dynamics of a probe massless particle confined by the harmonic potential near the horizon of the dyonic $\rm{AdS_4}$-Reissner-Nordstr\"om black hole. The total energy of the particle, chemical potential and magnetic field in this system serving as independently adjustable parameters tune nonlinearity and phase-space structure. By analyzing the trajectories on the Poincaré section and evaluating the Lyapunov exponents, we obtain the dynamical phase diagrams of the chaos and find their counteracting regulatory role: at low energy, chaos is enhanced and the Lyapunov exponent $\lambda_L$ violates its upper bound  (i.e. surface gravity) in the extremal black hole limit(combined paramete $\Gamma=3$); at high energy, the same extremal limit suppresses chaos, with $\lambda_L$ dropping to zero and a regular dynamical corridor emerging along $\Gamma=3$ in the dynamical phase diagrams. These results establish a direct mapping between black hole thermodynamics and microscopic chaos, offering new insights into the AdS/QCD correspondence and nonlinear dynamics in strongly curved spacetimes.}

\begin{document}
\maketitle
\flushbottom

\section{Introduction}
\label{sec:intro}

The exploration of particle dynamics around compact objects has long served as a cornerstone for probing the fundamental nature of gravity. Since the landmark detections of gravitational waves by LIGO \cite{LIGOScientific:2016aoc,LIGOScientific:2016sjg} and the direct imaging of event horizons by the Event Horizon Telescope \cite{EventHorizonTelescope:2019dse}, black holes have transitioned from theoretical constructs to accessible laboratories for testing General Relativity and its extensions. Among the myriad of phenomena occurring in these strongly curved spacetimes, the emergence of chaotic motion in probe particles has attracted significant attention, as it bridges the gap between macroscopic black hole thermodynamics and microscopic non-linear dynamics.

In the vicinity of a black hole horizon, the gravitational field is so intense that even simple integrable systems can exhibit complex, non-integrable behaviors when subjected to perturbations. Numerous works \cite{Dalui:2018qqv,Bera:2021lgw,Dalui:2025bwm,Liu:2025izb,Yang:2026gcv,An:2025xmb,Cao:2025qpz,Das:2024iuf,Gwak:2022xje,Shaymatov:2021qvt,Lei:2020clg,Takahashi:2008zh,Li:2018wtz} study the chaotic motion of particle near various black hole horizons, where the radial exponential instability of null geodesics serves as the primary driver for chaotic fluctuations. Most of these studies primarily focus on either various types of black holes or probe particles with different properties, and some interesting conclusions have been discovered: the parameters of the black hole itself or the external field around the black hole either enhance the chaos of particle motion from beginning to end under all conditions, or strengthen it under some conditions and weaken the chaotic behavior under other conditions. Unfortunately, these non-trivial behaviors have never been systematically analyzed and explained. For example, Ref.~\cite{Takahashi:2008zh} studies the chaotic motion of charged particles in an electromagnetic field surrounding a rotating black hole and finds that the dragging effects of the spacetime by a rotating black hole weaken the chaotic properties for some sets of initial parameters; Ref.~\cite{Li:2018wtz} studies the chaotic motion of neutral and charged particles in the magnetized
Ernst-Schwarzschild spacetime and shows that chaos is strengthened typically with an increase in the magnetic field under appropriate circumstances, while, for charged particles, the electromagnetic forces have an effect on strengthening or weakening the extent of chaos. A recent classic work~\cite{Hashimoto:2016dfz} establishes a profound connection between the event horizon and the onset of chaos. They demonstrate that the motion of a particle pulled by external forces near a spherically symmetric static black hole possesses a universal Lyapunov exponent $\lambda_L$, which is bounded by the surface gravity $\kappa$, i.e., $\lambda_L \leq \kappa$. This bound is remarkably identical to the Maldacena-Shenker-Stanford (MSS) bound on the rate of exponential growth of operator perturbations in quantum field theories~\cite{Maldacena:2015waa}, which was originally found by
thought experiments of shock waves near a black hole
horizon~\cite{Shenker:2013pqa,Shenker:2013yza} and the AdS/CFT correspondence~\cite{Cotler:2016fpe}. 

To comprehensively understand and analyze the non trivial conclusions mentioned above, we take a dyonic $\rm{AdS}_4$-Reissner-Nordström black hole as an example.  Although much of the existing work focuses on energy-driven transitions to chaos by considering the influence of different types of black holes, the role of electromagnetic charges—specifically in the context of dyonic black holes—remains relatively under-explored. Dyonic black holes, which carry both electric and magnetic charges, are of particular interest in the framework of the AdS/CFT correspondence~\cite{Rodrigues:2020ndy}. In the dual boundary field theory, the chemical potential $\mu$ and the background magnetic field $B$ represent independent control parameters that dictate the phase structure of strongly coupled systems. Previous investigations into magnetized Reissner-Nordström black holes have shown that the combined effect of magnetic fields and electric charges can significantly alter the innermost stable circular orbits (ISCO) and the center of mass energy extracted by collision of two particles~\cite{Shaymatov:2021qvt}. However, a systematic characterization of how these thermodynamic parameters modulate the phase-space topology of microscopic chaos near the horizon is still lacking.

In this work, we investigate the chaotic dynamics of a massless probe particle confined by external harmonic potentials near the horizon of a dyonic $\rm{AdS}_4$-Reissner-Nordström black hole. Our central contribution is the identification of a direct mapping between black hole thermodynamics and microscopic chaos. By treating $\mu$ and $B$ as independent control parameters, we move beyond the traditional energy-driven paradigm that treats energy as the sole driver of nonlinearity. By employing a combination of analytical near-horizon approximations and numerical techniques—including Poincaré sections, Maximum Lyapunov Exponents (MLEs), and the construction of detailed dynamical phase diagrams—we reveal a novel "counteracting regulation" mechanism. Specifically, we find that while an increase in either $\mu$ or $B$ generally promotes chaos in low-energy regimes, their interplay leads to a suppression of chaotic behavior in high-energy regimes. As the system approaches to the extremal black hole limit ($\Gamma \to 3$), this suppression manifests as a "corridor of regular dynamics" near the extremal curve, where the vanishing Hawking temperature quenches the horizon-induced instability.

Our results establish a direct mapping between the thermodynamic phase transitions of dyonic black holes and the transition between regularity and chaos in microscopic orbits. This not only offers new insights into the non-linear dynamics of strongly curved spacetimes but also provides a potential diagnostic tool for probing the QCD phase boundary~\cite{Zhao:2022uxc,Zhao:2023gur,Chen:2025goz,Zhu:2025gxo,Tang:2025llq,Zeng:2025tcz,Li:2025lmp,Cai:2022omk,Zhu:2025pxh,Liu:2024efy,Bu:2024fhz,Cao:2024jgt,Farias:2025vss,Yang:2023gfy} and the properties of heavy vector mesons through the AdS/QCD correspondence~\cite{Shukla:2024qlf,Shukla:2024wsu,PandoZayas:2010xpn,Hashimoto:2018fkb,Colangelo:2020tpr,Colangelo:2021kmn,Li:2025sje,Li:2024iuf,Li:2024yma,Hashimoto:2016wme}.

The remainder of this paper is organized as follows: In section~\ref{sec:metric}, we derive the equations of motion of a particle pulled by an scalar force, near a dyonic black hole horizon. In section~\ref{sec:toy}, we analyze a near-horizon radial toy model to build physical intuition. Section~\ref{sec:theory} discusses how $\mu$ and $B$ as control parameters affect the nonlinearity of the system. Section~\ref{sec:num} presents the numerical results and dynamical phase diagrams. Finally, we conclude with a summary and discussion of the holographic implications in section~\ref{sec:sum and dis}.

\section{Dynamics of particle around a dyonic \texorpdfstring{$\bf{AdS_4}$}{AdS4}-Reissner-Nordstr\"{o}m black hole}\label{sec:metric}

The dyonic  $\rm{AdS}_4$-Reissner-Nordstr\"{o}m black hole is a magnetically and electrically charged black brane solution to the Einstein–Maxwell equations with a negative cosmological constant, which is asymptotically anti-de Sitter in four dimensions. In Poincar\'{e} coordinates, its metric can be written as~\cite{Hartnoll:2009sz}
\begin{equation}\label{metric}
ds^2 = \frac{1}{z^2}\left(-f(z)dt^2 + \frac{dz^2}{f(z)} + dx^2 + dy^2\right),
\end{equation}
where $z=0$ corresponds to the $\rm{AdS}$ boundary, and the blackening factor $f(z)$ takes the form
\begin{equation}\label{blacken factor}
f(z) = 1 - \left(1+\mu^2 z_h^2+B^2z_h^4\right)\left(\frac{z}{z_h}\right)^3 + (\mu^2 z_h^2+B^2z_h^4)\left(\frac{z}{z_h}\right)^4,
\end{equation}
with $\mu$ and $B$ representing the chemical potential and magnetic field, respectively, and $z_h$ the horizon radius. The corresponding electromagnetic field strength $F$ is given by
\begin{equation}
F =dA= -\frac{\mu}{z_h} dz \wedge dt + B dx \wedge dy.
\end{equation}
The Hawking temperature of the black brane is determined by its surface gravity at the horizon. For a static metric of the form \eqref{metric}, the surface gravity can be computed via the standard formula $
\kappa = \lim_{z \to z_h} \frac{1}{2} \sqrt{\frac{g^{zz}}{-g^{tt}}} \left| \partial_z g_{tt} \right|.$
Substituting $g_{tt} = -\frac{1}{z^2}f(z)$ and $g^{zz} = z^2 f(z)$, and noting that $f(z_h)=0$, we obtain the compact result
\begin{equation}
\kappa = \frac{1}{2} |f'(z_h)|.
\end{equation}
Since $f(z)$ decreases from unity at the boundary to zero at the horizon, we have $f'(z_h) < 0$. Consequently, the Hawking temperature  of the dyonic black brane is then given by
\begin{equation}
T = \frac{\kappa}{2\pi} = -\frac{f'(z_h)}{4\pi}=\frac{1}{4\pi z_h} \left( 3 -z_h^2 (\mu^2 + z_h^2 B^2) \right).
\end{equation}
This expression reduces to the familiar result $T = 3/(4\pi z_h)$ for the neutral black brane ($\mu = B = 0$). The presence of electric and magnetic charges ($\mu$ and $B$) lowers the temperature.

The timelike Killing vector $\chi^\mu=(1,0,0,0)$ associated with the metric \eqref{metric} yields a conserved energy for geodesic motion, defined as $E=-\chi^\mu p_\mu=-p_t$, where $p_\mu=(p_t,p_z,p_x,p_y)$ represents the four-momentum vector. Expressing this quantity in terms of the spatial momentum components requires solving the mass-shell condition $g^{\mu\nu}p_\mu p_\nu=-m^2$. In the present study we restrict attention to massless particles($m=0$) propagating in the $z-x$ plane near the horizon. Substituting the explicit form of the metric leads to the dispersion relation
\begin{equation}
    E=\pm\sqrt{f(z)\left(f(z)p_z^2+p_x^2\right)},
\end{equation}
where the upper "$+$" denotes the energy for the outgoing particle, while, the lower "$-$" represents ingoing particle. We shall concentrate on outgoing modes, and therefore retain only the positive root in what follows.

To study bounded motion that does not fall into the horizon, we impose external harmonic confinements in the radial and transverse directions~\cite{Hashimoto:2016dfz,Dalui:2018qqv},
\begin{equation}
V_{\mathrm{harm}} = \frac{K_z}{2}(z-z_c)^2 + \frac{K_x}{2}(x-x_c)^2 ,
\end{equation}
with spring constants $K_z$, $K_x$ and equilibrium positions $z_c$, $x_c$. The full Hamiltonian for an outgoing particle then reads
\begin{equation}\label{eq:Hami}
    E=\sqrt{f(z)\left(f(z)p_z^2+p_x^2\right)}+\frac{K_z}{2}(z-z_c)^2+\frac{K_x}{2}(x-x_c)^2.
\end{equation}
In writing above Hamiltonian, we have neglected subleading cross‑couplings between the harmonic potentials and the curved background; This approximation is valid because the harmonic potentials are chosen to be strong, ensuring that the confining forces dominate over any curvature-induced corrections near the equilibrium positions. The resulting equations of motion follow from Hamilton’s equations,
\begin{align}
        \dot{z}&=\frac{\partial E}{\partial p_z}=\frac{p_zf(z)^2}{\sqrt{f(z)\left(f(z)p_z^2+p_x^2\right)}},\label{eq:eom1}\\
        \dot{p_z}&=-\frac{\partial E}{\partial z}=K_z\left(z_c-z\right) -\frac{\left(2f(z)p_z^2+p_x^2\right) f'(z)}{2\sqrt{f(z)\left(f(z)p_z^2+p_x^2\right)}},\label{eq:eom2}\\
        \dot{x}&=\frac{\partial E}{\partial p_x}=\frac{p_xf(z)}{\sqrt{f(z)\left(f(z)p_z^2+p_x^2\right)}},\label{eq:eom3}\\
        \dot{p_x}&=-\frac{\partial E}{\partial x}=K_x\left(x_c-x\right).\label{eq:eom4} 
\end{align}
Here, the "$^\cdot$" denotes differentiation with respect to the time $t$ along the particle’s worldline, i.e. $^\cdot=\frac{d}{dt}$.
\section{Analysis of a Near-Horizon Radial Toy Model}\label{sec:toy}

To understand the most fundamental influence of the black hole horizon on particle motion, we first analyze a highly simplified scenario: we neglect the particle's transverse motion ($p_x = 0$) and the external harmonic potential, focusing solely on its purely radial motion in the near-horizon region. Although this “toy model” is insufficient to describe genuine chaos, it cleanly reveals the physical essence of the horizon as a source of linear instability and provides crucial intuition for understanding the role of parameters $\mu$ and $B$. The corresponding trajectory equation of particles is simplified as
\begin{equation}
    \dot{z}=\frac{\partial E}{\partial p_z}=f(z),
\end{equation}
To simplify the analysis, we introduce a dimensionless combined parameter, $\Gamma \equiv \mu^2 z_h^2 + B^2 z_h^4$. The specific form of the blackening factor is

\begin{equation}
    f(z)=1-(1+\Gamma)\left(\frac{z}{z_h}\right)^{3}+\Gamma\left(\frac{z}{z_h}\right)^{4}.
\end{equation}
For nonextremal black hole ($T>0$), near-horizon blackening function can be approximated as
\begin{equation}
    f(z)\approx -\frac{(3 - \Gamma)}{z_h}(z - z_h)+O\left((z-z_h)^2\right).
\end{equation}
The physical meaning of \(\Gamma\) is clear: it comprehensively measures the strength of the electromagnetic charges (chemical potential and magnetic field) carried by the black hole in modifying the spacetime geometry. \(\Gamma = 0\) corresponds to an uncharged Schwarzschild-AdS black hole. The corresponding trajectory equation is
\begin{equation}
    \dot{z}\simeq\frac{\Gamma-3}{z_h}(z-z_h)+O\left((z-z_h)^2\right).
\end{equation}
The above equation has an analytical solution:
\begin{equation}
    z=z_h+c_1 e^{\frac{\lambda  \left(\Gamma-3\right)}{z_h}},
\end{equation}
where $c_1$ is the integration constant. The exponential growth of the radial coordinate signifies a fundamental linear instability induced by the horizon. In itself, this instability does not constitute chaos — for genuine chaotic dynamics, at least two nonlinearly coupled degrees of freedom are required — but it provides the essential driving mechanism. In the full dynamical system, where the particle possesses both radial and transverse motion (\(p_x \neq 0\)) and is confined within a harmonic potential (\(K_z, K_x \neq 0\)), this radial instability couples nonlinearly to the transverse degree of freedom. Together with the geometric nonlinearity encoded in $f(z)$, it can break the integrability of the harmonic system and drive a transition from regular to chaotic motion.

However, if we take $\Gamma=3$, which corresponds to extremal dyonic black holes ($T=0$), one clearly finds that the trajectory reduces to $z = z_h + c_1$. In this extremal limit, the linear radial instability vanishes, and with it the primary engine for chaos is removed. The subsequent analysis of the full system will confirm that the onset of chaos is strongly suppressed near the extremal black hole limit.

\section{Analysis of Magnetic Field and Chemical Potential as Control Parameters for Chaos}
\label{sec:theory}

In section \ref{sec:metric}, we derived the complete Hamiltonian and the equations of motion for a probe particle in the black hole background under the constraint of harmonic potentials. The purpose of this section is to theoretically elucidate how, in addition to the conserved energy of the particle, the chemical potential and the magnetic field of the black hole act as independent and crucial control parameters that profoundly influence the nonlinear strength and the phase-space structure of the system.

\subsection{Structure of the Hamiltonian System and Sources of Nonlinearity}
\label{subsec:hamiltonian_structure}

The system under study is described by the Hamiltonian \eqref{eq:Hami}, where \(f(z) \equiv f(z; \mu, B)\) is the blackening factor carrying parameter dependencies. The relativistic gravitational term, $\sqrt{ f(z) [ f(z) p_z^2 + p_x^2 ] }$, introduces strong nonlinear coupling between the coordinates and momenta. Here, the blackening factor \(f(z)\) not only acts as a potential barrier but also couples with the momenta \((p_z, p_x)\) in a non-trivial way, breaking the additive separation of kinetic and potential energy typical of classical mechanics. This is the geometric source of complex dynamics in relativistic systems. The harmonic potentials $V_{\mathrm{harm}}$ are quadratic and therefore yield linear restoring forces. Their role is to confine the particle motion to a finite region near the horizon; they do not by themselves generate nonlinearity. However, they shape the overall phase‑space structure and, together with the gravitational nonlinearity, determine the onset of chaotic behavior. The parameters $\mu$ and $B$ modify the shape of $f(z)$ and hence dictate how the curved spacetime couples with the harmonic oscillations. Therefore, we expect $\mu$ and $B$ to act as effective control parameters for the system’s nonlinearity and chaotic behavior, although their influence may be modulated by the strong constraint(conserved energy).

\subsection{Analysis of the Parameter Dependence of the Blackening Factor \texorpdfstring{$f(z; \mu, B)$}{f(z;mu,B)}}
\label{subsec:blackening_factor}

\begin{figure}
    \centering
    \includegraphics[width=0.8\linewidth]{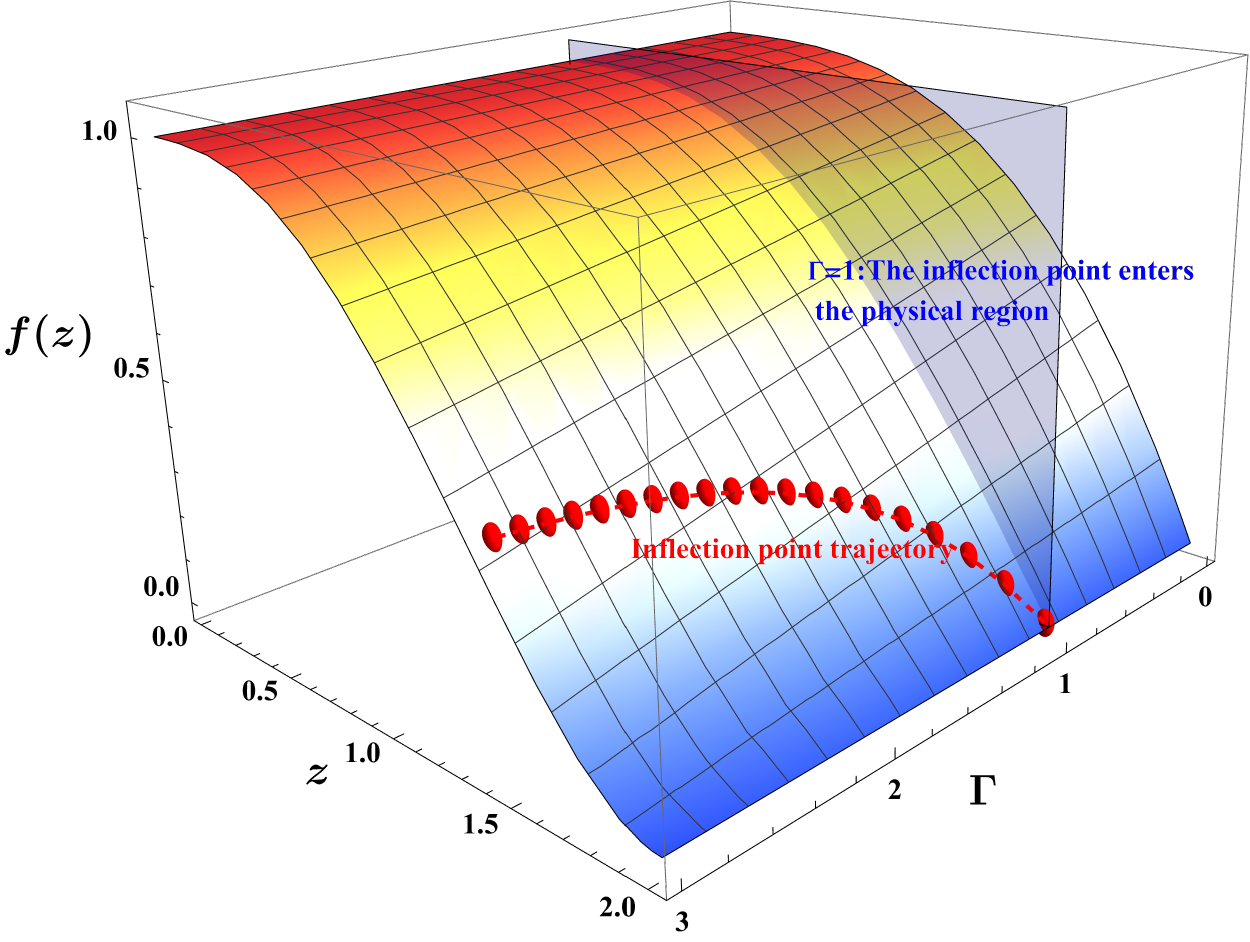}
    \caption{Evolution of the blackening factor \( f(z) \) with respect to the parameter 
        \(\Gamma = \mu^2 z_h^2 + B^2 z_h^4\). 
        The surface shows the value of \( f(z) \), while the red dashed trajectory marks 
        the position of the inflection point \((f''(z) = 0)\). 
        When \(\Gamma < 1\), the inflection point lies outside the physical region 
        \((0 < z < z_h)\); when \(\Gamma > 1\), it enters the physical region, 
        indicating a pronounced distortion in the functional shape. 
        The blue plane marks the critical value \(\Gamma = 1\), at which the inflection 
        point is located precisely at the horizon \(z = z_h\).}
    \label{fig:f}
\end{figure}

An increase in \(\Gamma\) alters \(f(z)\) in two ways: (1) Change in near-horizon slope: Expanding around the horizon \(z=z_h\), we have \(f'(z_h) = -(3-\Gamma)/z_h\), which determines the strength of the ``surface gravity'' near the horizon. When \(\Gamma < 3\), increasing \(\Gamma\) decreases \(|f'(z_h)|\), meaning the gravitational gradient near the horizon becomes milder. (2) Distortion of the overall functional form: An increase in \(\Gamma\) enhances the coefficient of the \((\frac{z}{z_h})^4\) term. This higher-order term can cause \(f(z)\) to develop inflection points, i.e., \(z=\frac{z_h(1+\Gamma)}{2\Gamma}\) where \(f''(z)=0\), in the region \(z < z_h\) (the main region of particle motion). Consequently, \(f(z)\) is no longer a simple monotonic function but possess a more complex convex-concave structure.  As shown in Fig.~\ref{fig:f}, the geometric shape of the blackening factor \( f(z) \) depends strongly on the combined parameter \(\Gamma\). For \(\Gamma > 1\), an inflection point (red trajectory) appears within the physical region. In the equations of motion~\eqref{eq:eom1}-\eqref{eq:eom4}, \(f(z)\) and its derivative \(f'(z)\) are nonlinearly coupled with the momenta. Therefore, the increased complexity of \(f(z)\)'s shape directly enhances the nonlinearity of the entire system of differential equations, indicating how \(\mu\) and \(B\) promote chaotic dynamics. 

Therefore, we can take it for granted that, for a fixed conserved energy, increasing the chemical potential or the magnetic field (i.e. increasing \(\Gamma\)) will significantly enhance the overall nonlinearity of the blackening factor \(f(z)\). This is expected to lower the critical energy threshold for the onset of chaos in the system, causing the Kolmogorov-Arnold-Moser (KAM) invariant tori to break at lower energies, and leading to the earlier and more widespread formation of chaotic regions in phase space.

\subsection{The Extremal Black Hole Limit: Qualitative Change in Dynamics}\label{subsec:extremal_limit}

When the parameters satisfy \(\Gamma = 3\), the system reaches the extremal black hole limit corresponding to the emergence of the $\rm{AdS}_2 \times \mathbb{R}^2$ space. At this point, the near-horizon expansion undergoes a fundamental change:
\begin{equation}
    f(z)|_{T=0}=1-4\left(\frac{z}{z_h}\right)^3+3\left(\frac{z}{z_h}\right)^4\approx \frac{6}{z_h^2}(z-z_h)^2+ \mathcal{O}((z-z_h)^3).
\end{equation}
Here, \(f(z)\) exhibits a double zero at \(z_h\). This change in geometric feature has profound implications for the dynamics. In the non-extremal case, the near-horizon motion \(\dot{z} \propto f(z) \propto (z-z_h)\) leads to exponential divergence or convergence of the radial coordinate, which is a typical source of local instability. In contrast, in the extremal case, \(\dot{z} \propto (z-z_h)^2\), resulting in a power-law behavior. This "softened" boundary condition greatly suppresses the exponential instability near the horizon.

Therefore, we expect that, when the system parameters \((\mu, B)\) are tuned to approach to the extremal limit (\(\Gamma \to 3\)), the strong source of chaos induced by the horizon is weakened. We predict that at a higher fixed energy, the chaotic motion observed under non-extremal parameters will be suppressed, and regions of regular motion (KAM tori) in phase space will reappear or significantly expand. This signifies a drastic change in dynamical behavior induced by the thermodynamic phase transition of the black hole.

\section{Numerical analysis}\label{sec:num}
In this section, we will intuitively reveal the overall phase-space structure of particle motion near the black hole horizon. To qualitatively distinguish between regular and chaotic motion, we first employed the Poincar\'{e} section method. For a constrained conservative Hamiltonian system, its Poincar\'{e} section reduces the continuous phase-space flow to a discrete map of intersection points. If the particle's motion is regular (e.g., periodic or quasi-periodic orbits), the corresponding points will form smooth closed curves or discrete fixed points on the section. Conversely, if the motion is chaotic, these points will exhibit seemingly random, scattered distributions within a finite area. Although the Poincar\'{e} section provides powerful qualitative imagery, to achieve a quantitative and objective characterization of the chaotic dynamics and to verify the inferences drawn from the section plots, we further computed the Maximum Lyapunov Exponents (MLEs) for the orbits. This exponent quantitatively measures the average exponential divergence rate of infinitesimally nearby trajectories: a value converging to zero corresponds to regular motion, while convergence to a positive constant confirms the presence of chaotic motion. In the subsequent numerical calculations, we take $z_h=2, z_c=1.2, x_c=0, K_z=K_x=50$, which corresponds to the bound of surface gravity $\kappa=\frac{3-\Gamma}{2z_h}=\frac{3-\Gamma}{4}$. 

It is noteworthy that from the numerical results of the figure~\ref{fig:mle} (Lyapunov exponent analysis) and figure~\ref{fig:DPHD} (dynamical phase diagram), we observe an energy‑dependent behavioural pattern: when the particle energy is low and its motion lies farther from the black‑hole horizon(as shown in figure~\ref{fig:E}), even near the extremal black hole ($\Gamma=3$) (i.e. the surface gravity $\kappa$ approaches to zero) the Lyapunov exponent $\lambda_L$ can remain to be a finite positive value, which clearly violates the upper bound $\lambda_L\leq \kappa$. This indicates that at low energy the electromagnetic charges ($\mu$ and $B$) are able to sustain chaotic dynamics even near zero temperature by enhancing the geometric nonlinearity. On the other hand, when the particle energy is high and its motion approaches to the horizon, the Lyapunov exponent rapidly decays to nearly zero in the vicinity of the extremal curve, thereby restoring and satisfying the bound $\lambda_L\leq \kappa$. This corresponds to the emergence of the “regular‑dynamics corridor” along the extremal curve in the phase diagram. As analyzed in section \ref{sec:theory}: the energy‑dependent geometric location leads to the violation and restoration of the Lyapunov bound—at low energy the particle stays farther from the horizon and chaos is mainly driven by the electromagnetic‑charge‑enhanced nonlinearity; at high energy the particle moves close to the horizon and the dynamics is dominated by the “softening” effect of the extremal black hole, which suppresses chaos.

\subsection{Poincar\'{e} sections}

\begin{figure}[tbp]
\centering
\includegraphics[width=.45\textwidth]{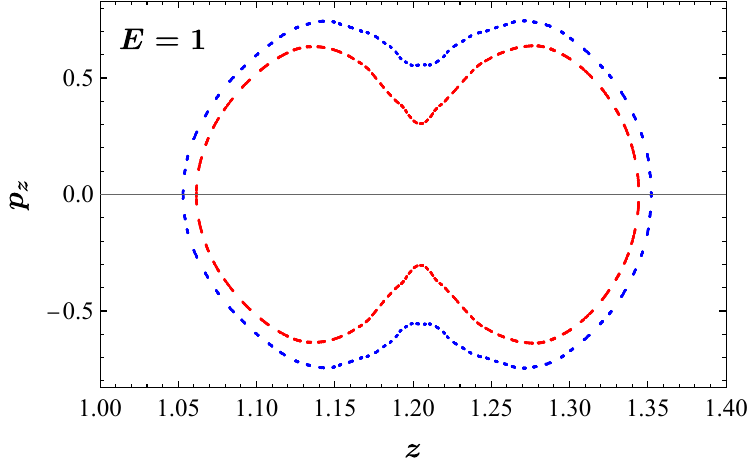}
\qquad
\includegraphics[width=.45\textwidth]{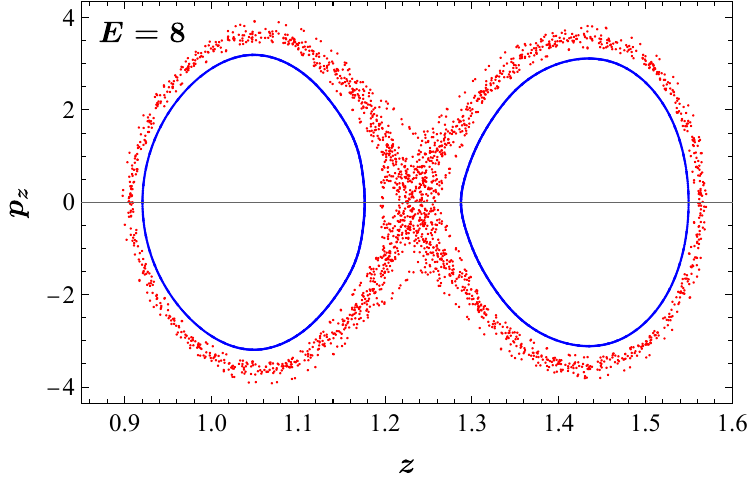}\\[0.7cm] % <--- 在这里修改，增加 0.7cm 的间距
\includegraphics[width=.45\textwidth]{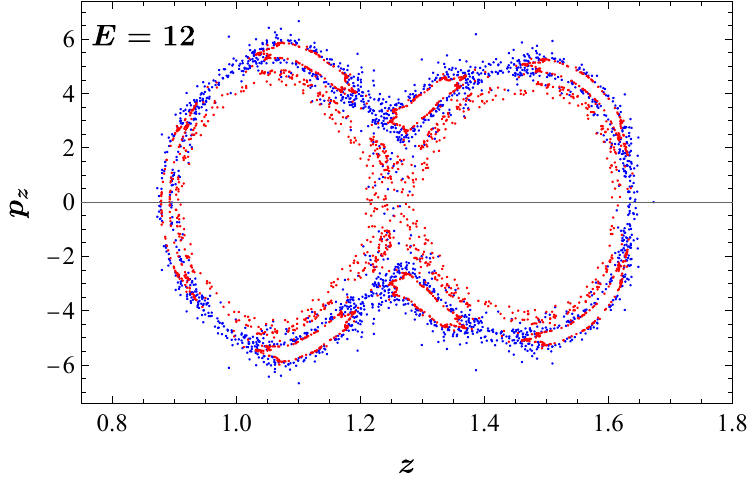}
\qquad
\includegraphics[width=.45\textwidth]{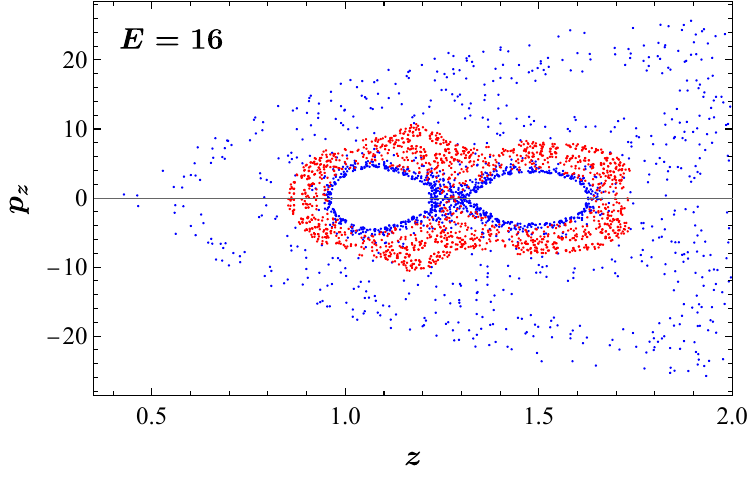}
\caption{The Poincar\'{e} sections of the particle motion near the black hole horizon for different energies with $\mu=B=0$. As the energy increases, the Kolmogorov-Arnold-Moser (KAM) tori tend to break.\label{fig:E}}
\end{figure}

In this subsection, we use Poincar\'{e} sections to systematically portray the overall evolution of massless probe-particle orbital structures near the black hole horizon as the energy, magnetic field and chemical potential vary. For clarity, we only consider two random initial conditions: 1. $z=1.2948, p_z=0.7001$; 2. $z=1.2741, p_z=0.6391$. The section is given by the condition $p_x>0$ and $x=0$. For different initial conditions, the specific value of transverse momentum $p_x$ is determined by the conservation energy.

\begin{figure}[tbp]
\centering
\includegraphics[width=.45\textwidth]{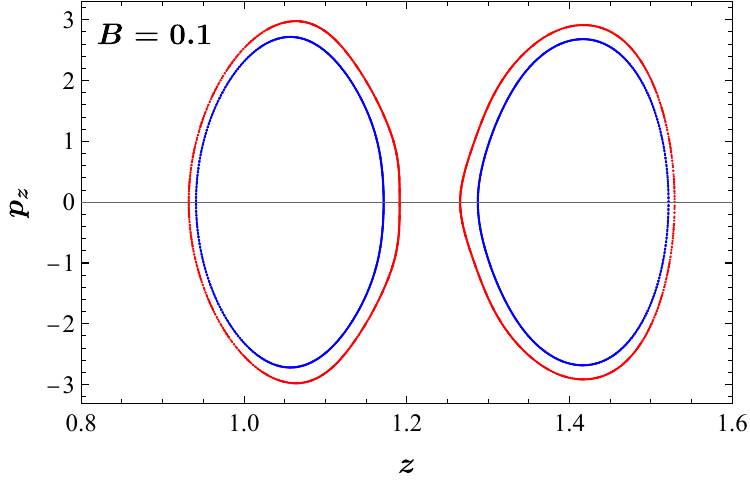}
\qquad
\includegraphics[width=.45\textwidth]{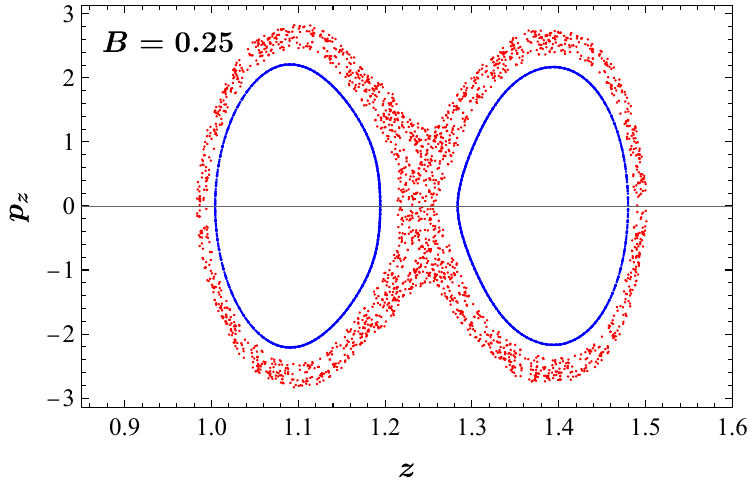}\\%[0.7cm]  <--- 在这里修改，增加 0.7cm 的间距
\includegraphics[width=.45\textwidth]{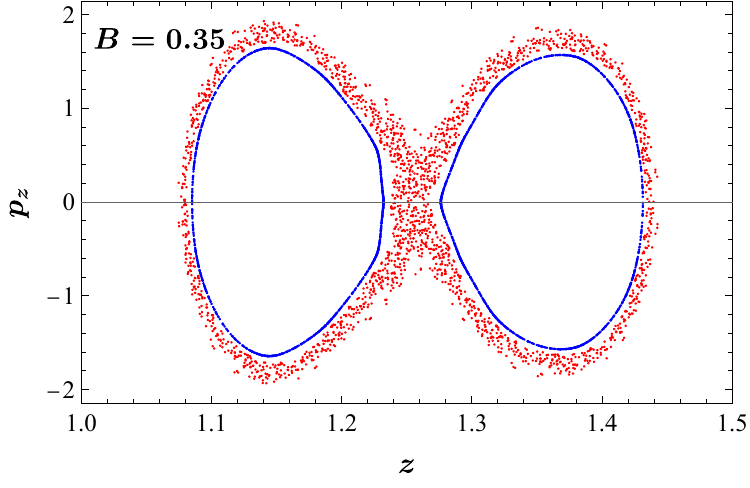}
\qquad
\includegraphics[width=.45\textwidth]{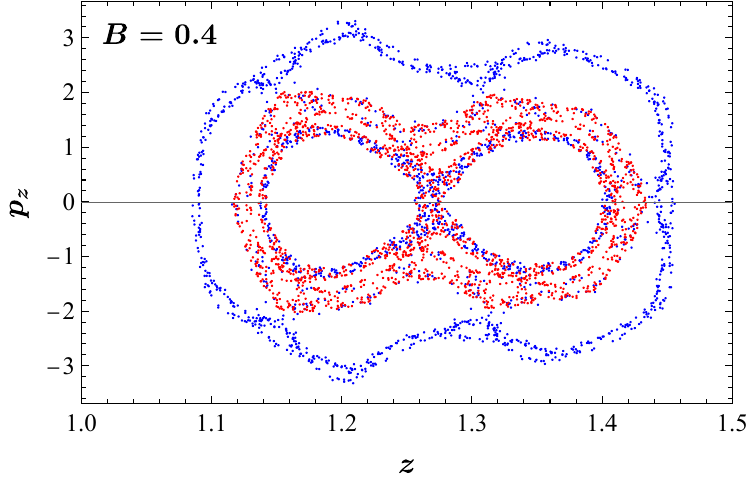}
\caption{The Poincar\'{e} sections of the particle motion near the black hole horizon for different magnetic fields with $\mu=0, E=8$. As the magnetic field increases, the Kolmogorov-Arnold-Moser (KAM) tori tend to break.\label{fig:B}}
\end{figure}

To expose the geometric and dynamical essence of particle motion in a purely gravitational field, we first analyze the Poincar\'{e} sections as a function of energy under ideal conditions with zero magnetic field and zero chemical potential, as shown in figure \ref{fig:E}. The green and orange points correspond to phase-space intersections from initial conditions 1 and 2, respectively. At lower energies $E=1$ (the upper-left panel), the particle trajectory is relatively far from the black hole's event horizon, so the section shows regular, nested closed curves, corresponding to typical Kolmogorov–Arnold–Moser (KAM) tori, which represent regular orbits in integrable or nearly integrable motion. As the particle energy is increased, the trajectory of particle motion gradually approaches to the black hole horizon, so the originally smooth and continuous tori gradually distort and twist, and ultimately break up and vanish, being replaced by progressively larger and more complex chaotic layers (scattered-point regions), as shown in the upper-right and lower panels of figure \ref{fig:E}. Note that the energy has an upper bound, because a particle with too much energy will fall into the black hole horizon, thus chaos is not well defined and the numerical calculation becomes extremely unstable. This evolutionary process clearly illustrates the torus-breakup mechanism predicted by the renowned KAM theory in nonlinear Hamiltonian systems\footnote{With the strengthening of the effective system perturbation—here equivalently induced by the enhanced nonlinearity due to increasing energy—more and more invariant tori are destroyed, and the regular motion regions of the system are gradually eroded by chaotic dynamics.}. 

\begin{figure}[tbp]
\centering
\includegraphics[width=.45\textwidth]{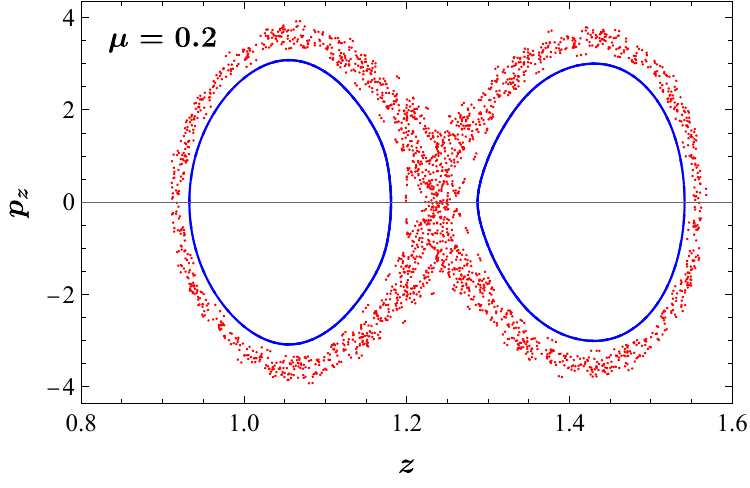}
\qquad
\includegraphics[width=.45\textwidth]{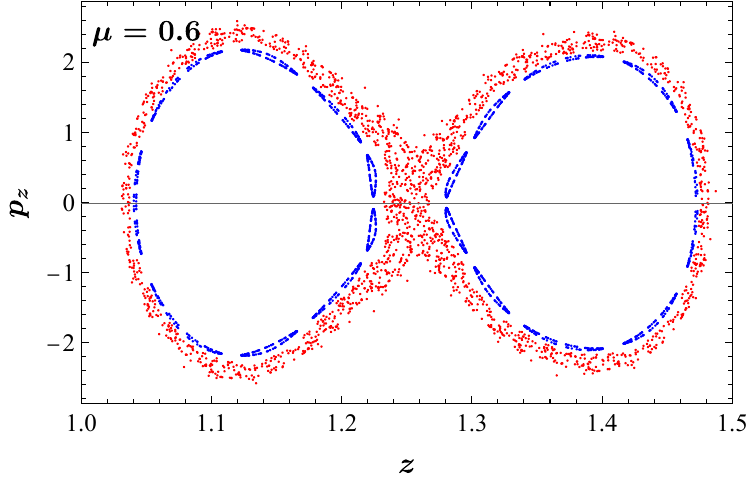}\\%[0.7cm]  <--- 在这里修改，增加 0.7cm 的间距
\includegraphics[width=.45\textwidth]{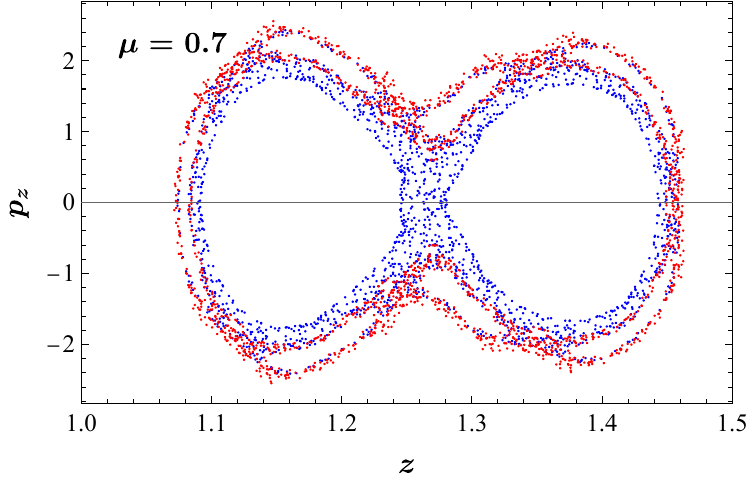}
\qquad
\includegraphics[width=.45\textwidth]{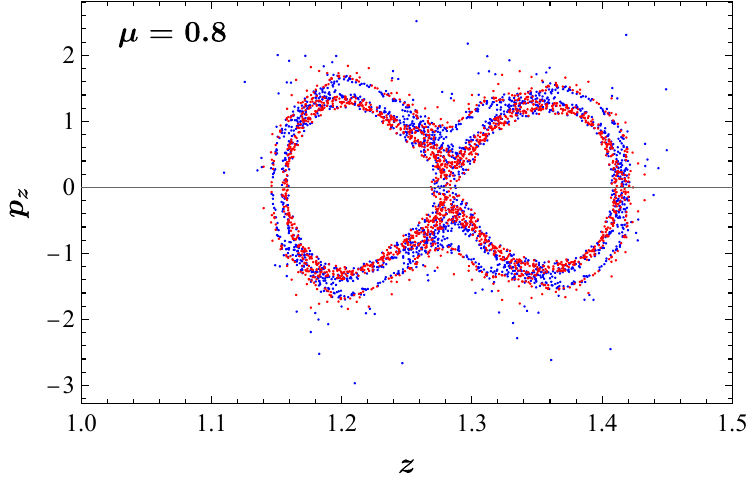}
\caption{The Poincar\'{e} sections of the particle motion near the black hole horizon for different chemical potentials with $B=0, E=7$. As the chemical potential increases, the Kolmogorov-Arnold-Moser (KAM) tori tend to break.\label{fig:mu}}
\end{figure}

Figures~\ref{fig:B} and~\ref{fig:mu} present a comparative view of how the black hole's electromagnetic charges independently influence the phase-space structure. Figure~\ref{fig:B} shows Poincar\'{e} sections for varying magnetic field $B$ with $\mu=0$ and energy $E=8$, while figure~\ref{fig:mu} shows sections  for varying chemical potential $\mu$ with $B=0$ and energy $E=7$. In both cases, an increase in the respective parameter raises the combined geometric factor $\Gamma = \mu^2 z_h^2 + B^2 z_h^4$, which enhances the nonlinear distortion of the blackening factor $f(z)$. This distortion strengthens the coupling between radial and transverse degrees of freedom, leading to the progressive breakup of Kolmogorov–Arnold–Moser (KAM) tori and the expansion of chaotic layers. The visual similarity between the two sequences confirms that both $\mu$ and $B$ act as effective, non-thermal control parameters, modulating horizon-induced chaos through the same geometric mechanism embodied in $\Gamma$. This parallel behavior underscores the multi-parameter controllability of the system's nonlinear dynamics.

\subsection{Lyapunov exponents}

\begin{figure}[htbp]
\centering
\includegraphics[width=.45\textwidth]{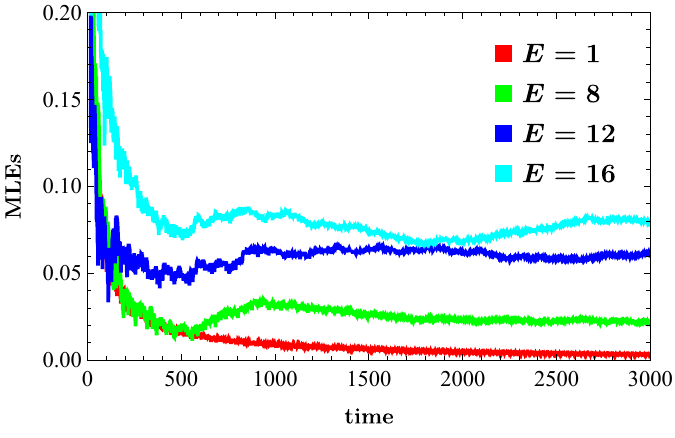}
\qquad
\includegraphics[width=.45\textwidth]{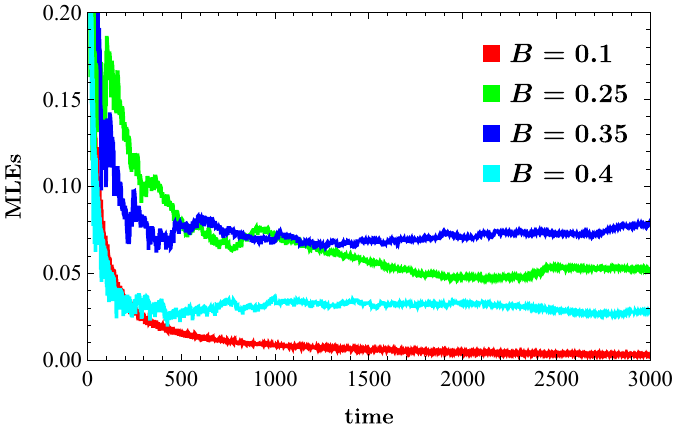}
\qquad
\includegraphics[width=.45\textwidth]{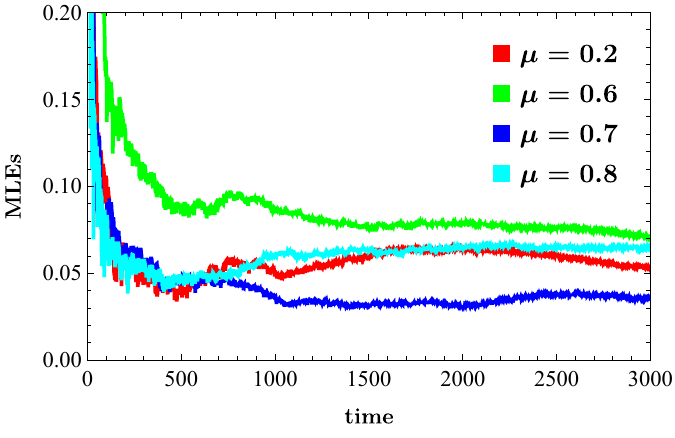}
\caption{Numerical analysis of the maximum Lyapunov exponents (MLEs). After a period of time, the MLEs converge to a set of values.\label{fig:mle}}
\end{figure}

In this subsection, to quantitatively measure the "scattered point clouds" observed in the Poincar\'{e} sections, avoiding potential subjective visual misinterpretation (e.g., mistaking a very dense, not-yet-fully-filled quasi-periodic orbit for chaos), we will compute the maximum Lyapunov exponents of the confined probe-particle system using the numerical method of Ref.~\cite{1996Numerical} (See the appendix~\ref{asec:lyapunov} for more details). The specific value of the maximum Lyapunov exponent can not only quantify the strength of chaos, but also be used to investigate how it changes systematically with physical parameters. Combining the global geometric view from the Poincar\'{e} section with the local quantitative analysis from the maximum Lyapunov exponent constitutes a mutually verifying, comprehensive analytical framework. This ensures the reliability and persuasiveness of our conclusion regarding "black hole horizon-induced chaos". 

Figure \ref{fig:mle} presents the numerical convergence of the maximum Lyapunov exponents (MLEs) as a function of time, computed for a representative trajectory near the black hole horizon (we use initial condition 2). After a transient period, the MLEs converges to a stable positive value for chaotic orbits, indicating exponential divergence of nearby trajectories, whereas for regular orbits it decays to zero. It can be observed that as the energy increases, the MLE gradually increases. However, at a fixed moderate energy, the MLE varies non-monotonically with both the magnetic field and the chemical potential—a behavior in full agreement with the theoretical analysis presented in section \ref{sec:theory}.

\subsection{Dynamical Phase Diagram}
\label{subsec:phase_diagram}

\begin{figure}[tbp]
\centering
\includegraphics[width=.45\textwidth]{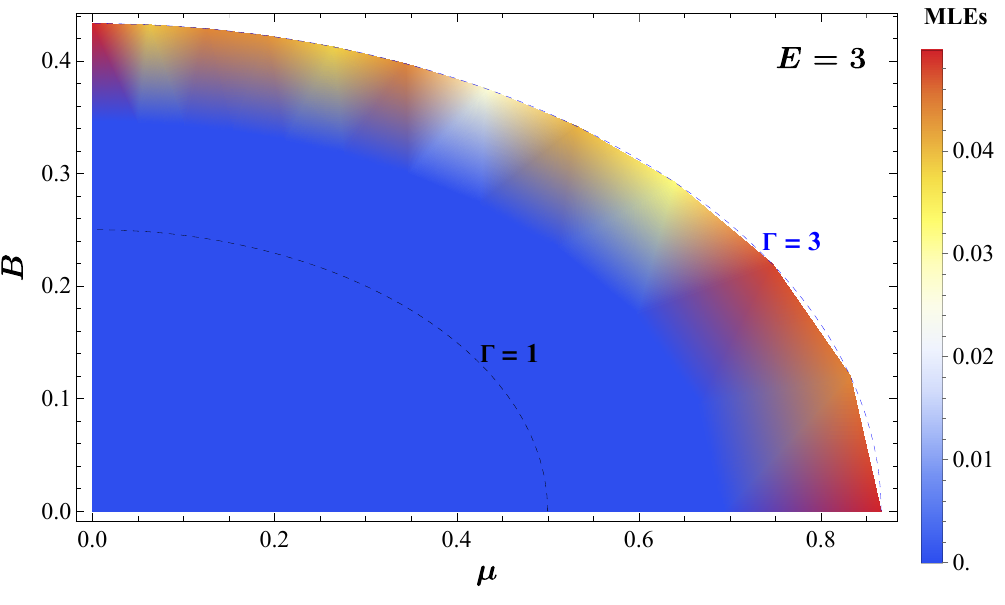}
\qquad
\includegraphics[width=.45\textwidth]{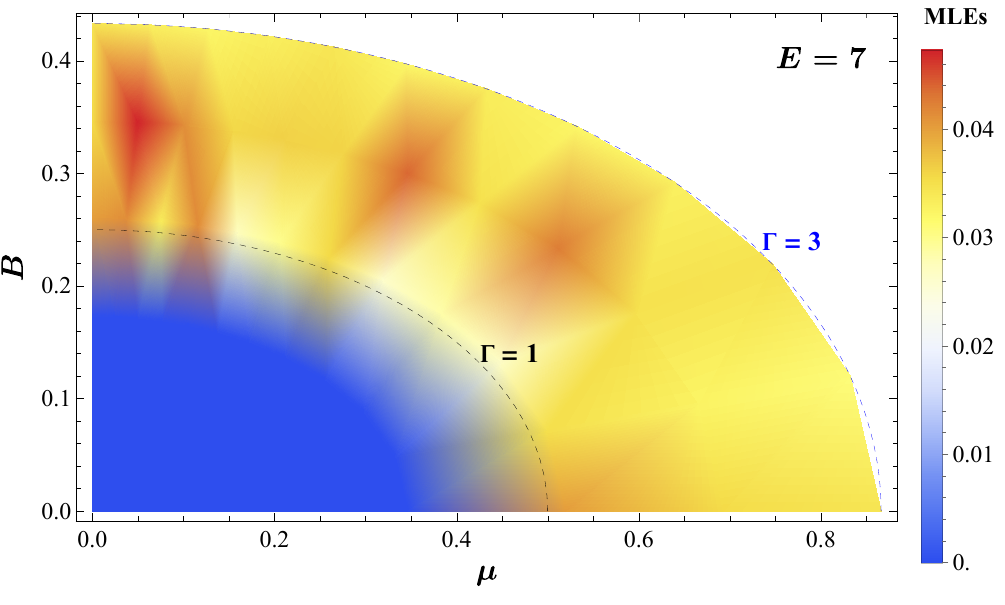}\\[0.7cm] % <--- 在这里修改，增加 0.7cm 的间距
\includegraphics[width=.45\textwidth]{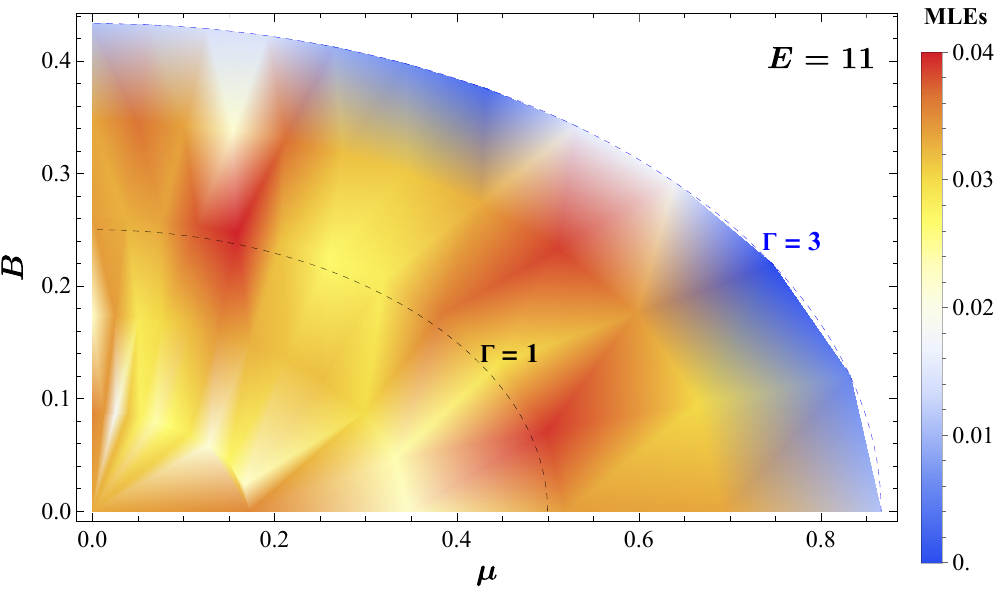}
\qquad
\includegraphics[width=.45\textwidth]{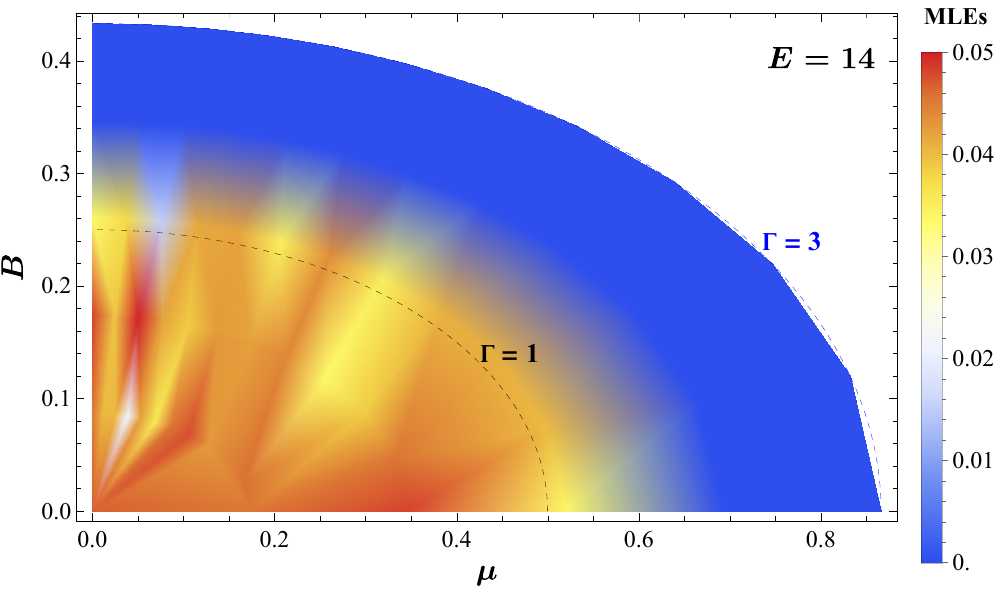}
\caption{Dynamical phase diagrams in the $(\mu, B)$ plane for four representative energy values: $E = 3, 7, 11, 14$. The color scale indicates the degree of chaoticity, quantified by the maximum Lyapunov exponent (MLE): blue regions correspond to regular motion (MLE $\approx 0$), red to strongly chaotic motion (MLE $> 0$). The dashed blue curve marks the extremal condition $\Gamma = \mu^2 z_h^2 + B^2 z_h^4 = 3$, along which a corridor of regular dynamics emerges for higher energy, separating broader chaotic domains. These diagrams visually demonstrate how the black hole's electromagnetic charges control the transition between regularity and chaos, and how the approach to extremality suppresses chaotic behavior at higher energies.\label{fig:DPHD}}
\end{figure}

Although in the static background solution, \(\mu\) and \(B\) contribute equally in defining \(\Gamma\), they originate from different electromagnetic field components and correspond to distinct physical perturbations in the boundary field theory. Therefore, they may have non-equivalent effects through other details not fully reduced in the equations of motion (e.g., subtle influences on boundary conditions). This could lead to slight anisotropy in the phase diagram along the \(\mu\) and \(B\) directions, meaning that the rate at which the system tends toward chaos may differ along the $\mu$ and $B$ axes. To globally understand the dynamical behavior in the \((\mu, B)\) parameter space, we will plot the dynamical phase diagram. This requires us to go beyond single-parameter scans and treat the \((\mu, B)\) plane as a control plane, systematically characterizing the dynamical phase (regular, chaotic, or mixed) at each point. 

Figure \ref{fig:DPHD} illustrates the dynamical phase diagram in the $(\mu, B)$ parameter space for four representative energy values $E = 3, 7, 11, 14$. Each subplot classifies the system's long-term behavior into regular, chaotic, or mixed phases based on the MLE and Poincar\'{e} section morphology. If the particle motion occurs in a region relatively far from the event horizon of a black hole(corresponding to lower energy), then near the origin of the dynamical phase diagram, where $\Gamma\approx0$, nonlinearity is weakest, and the system behavior is closest to integrability, with regular orbits dominating. As $\mu$ or $B$ increases, nonlinearity strengthens, and chaotic regions expand from local layers to a global sea in phase space. In the parameter plane, this manifests as a ``chaotic phase'' extending outward from the regular region. Significantly, even near the extreme black hole ($\Gamma=3$) where the surface gravity approaches to zero, the Lyapunov exponent can still remain to be a finite positive value, which clearly violates the upper bound $\lambda_L\leq \kappa$. However, if the particle motion occurs in a region very close to the event horizon of a black hole(corresponding to higher energy), then near the origin of the dynamical phase diagram, nonlinearity is strongest, and the system behavior is completely non-integrable, with chaotic sea dominating. As $\mu$ or $B$ increases, nonlinearity weakens, and chaotic regions shrink from the global sea to local layers in phase space. In the parameter plane, this manifests as a ''regular phase'' extending outward from the chaotic region. Notably, due to the dynamical quenching effect, a corridor of regular dynamics emerges along the extremal curve $\Gamma = 3$ near the black hole horizon, which restores and satisfies the upper bound $\lambda_L\leq \kappa$. This regular corridor separates broader chaotic regions, visually mapping the thermodynamic extremality condition onto a dynamical transition boundary. 

In summary, we can draw the following conclusion: the effects of the magnetic field and the chemical potential consistently exhibit a counteracting regulation on the system’s dynamics: if the system initially lies in a regular state, they drive it toward chaos; if it starts in a chaotic state, they tend to steer it back toward regular motion. This “counteracting regulation” becomes particularly pronounced under high-energy conditions very close to the black hole horizon, where it further gives rise to a corridor of regular dynamics near the extremal curve.

\section{Summary and discussion}\label{sec:sum and dis}

In this work, we have systematically explored how the electromagnetic charges of a dyonic AdS-Reissner-Nordstr\"om black hole---encoded in the chemical potential $\mu$ and magnetic field $B$---govern the onset and suppression of chaotic motion for a confined probe particle. By analyzing the near-horizon geometry through the blackening factor $f(z; \mu, B)$, we derived a Hamiltonian formulation and identified $\mu$ and $B$ as key parameters that tune nonlinear coupling between degrees of freedom. We use the Poincar\'e sections and Lyapunov exponents to anlyze the dynamical behavior. The former aims to globally and qualitatively reveal the evolution of the phase-space structure, while the latter is used to quantitatively confirm and characterize the chaotic behavior. The numerical diagnostics have confirmed that, in a region relatively far from the event horizon of a black hole(corresponding to lower energy), increasing $\mu$ or $B$ generally enhances chaos, fragmenting Kolmogorov-Arnold-Moser tori and expanding chaotic seas in phase space. Remarkably, even as the extremal limit ($\Gamma=3$) is approached—where the surface gravity vanishes—the Lyapunov exponent can remain finite and positive, thereby violating the upper bound $\lambda_L \leq \kappa$. This apparent violation may hint at a fundamental distinction between classical orbital instability near the horizon and quantum information scrambling~\cite{Maldacena:2015waa}. Strikingly, as the system approaches to the extremal black hole limit ($\Gamma = 3$), chaotic dynamics are quenched in a region very close to the event horizon of a black hole(corresponding to higher energy), giving rise to a corridor of regular motion in the $(\mu, B)$ plane. In this regime, $\lambda_L$ drops to nearly zero, satisfying the bound $\lambda_L \leq \kappa$ and confirming the suppression of chaos. This corridor aligns precisely with the thermodynamic extremality condition, thereby bridging macroscopic black hole phases with microscopic dynamical behavior. 

While the regular dynamical corridor is a theoretical construct under idealised conditions, its potential observational signatures warrant further exploration. In astrophysical settings where black holes possess substantial electromagnetic charges (e.g., magnetised black holes in active galactic nuclei) ~\cite{EventHorizonTelescope:2019dse,Urry:1995mg,Fermi-LAT:2009ihh,SDSS:2003asf}, the transition between regular and chaotic motion of surrounding matter could imprint on electromagnetic signals such as quasi‑periodic oscillations in X‑ray flux~\cite{Kaspi:2017fwg,Watts:2016uzu,Strohmayer:1996zz,Ingram:2019mna,Israel:2005av,VanderKlis:1996xn,Kolos:2017ojf,Mendez:1998eg} or coherent polarisation patterns~\cite{Stephenson:2020onf,Deshpande:2000ix}. The high‑precision gravitational‑wave detectors (e.g., LISA) may also discern the phase coherence of extreme mass‑ratio inspirals~\cite{Robson:2018ifk,Babak:2017tow,LISA:2022yao,Barack:2006pq,Kocsis:2011dr,Katz:2021yft,Cardoso:2022whc,Yunes:2011aa,Chua:2020stf,Moore:2016qxz} modulated by such regular corridors. On the experimental side, analogue gravity systems (e.g., acoustic black holes in Bose–Einstein condensates)~\cite{Barcelo:2005fc,MunozdeNova:2018fxv,Visser:2001fe,Fagnocchi:2010sn,Torres:2020tzs,Faccio:2013kpa} offer a controllable platform to simulate the effective geometry and test the parameter‑dependent suppression of chaos. Finally, through the gauge/gravity duality~\cite{Maldacena:1997re}, the regular corridor may correspond to a regime of slow thermalisation in the boundary quantum field theory~\cite{Horowitz:1999jd,Maldacena:2015waa, Michailidis:2019tgg}, accessible via out‑of‑time‑order correlators~\cite{Hashimoto:2017oit} or spectral functions~\cite{Hartnoll:2009sz}.

As is well known, particles exhibit chaotic behavior near the black hole horizon, while in AdS/QCD, the phase transition of hadronic matter corresponds to the large/small black hole phase transition~\cite{Bezboruah:2025udi, Zhang:2025cdx, Lei:2024qpu,Lyu:2023sih}. Therefore, we may potentially probe the QCD phase boundary by investigating the chaotic behavior of particles. Furthermore, the maximum Lyapunov exponent could be employed as an order parameter to study the properties of vector mesons~\cite{Ferreira:2025iqe,Grossi:2025hxg,Zhu:2024dwx,Wang:2024rim,Chen:2024edy}, such as their melting temperature and spin alignment~\cite{Sheng:2024kgg,Zhao:2024ipr,Yan:2025tlx,Sahoo:2025bkx,Zhu:2025rdj,Chen:2025mrf,Ahmed:2025bwi,Zhang:2024mhs,Yang:2024fkn,Wei:2024lah,Sun:2024anu}, which will be extremely interesting.

\appendix
\section{Calculation of Lyapunov Exponents for Classical Chaotic Systems}\label{asec:lyapunov}

To quantitatively characterize the strength of chaos in particle motion near the black hole horizon, we systematically compute the system's maximum Lyapunov exponent (MLE). This exponent is defined as the average exponential divergence rate of two infinitesimally close trajectories in phase space:
\begin{equation}
\lambda_{L} = \lim_{t \to \infty} \lim_{\|\delta \mathbf{X}_0\| \to 0} \frac{1}{t} \ln \frac{\|\delta \mathbf{X}(t)\|}{\|\delta \mathbf{X}_0\|},
\end{equation}
where \(\delta \mathbf{X}_0\) denotes the initial perturbation and \(\delta \mathbf{X}(t)\) the perturbation after evolution time \(t\). If \(\lambda_{L} > 0\), the system exhibits chaotic motion; if \(\lambda_{L} = 0\), the motion is regular.

In actual numerical computation, directly integrating two infinitely close orbits is not feasible because numerical errors would rapidly amplify. We therefore adopt the standard algorithm based on variational equations~\cite{1989Practical} and Gram--Schmidt orthogonalization~\cite{article}, which stably extracts the full Lyapunov spectrum by linearizing the evolution along the reference orbit.

\subsection{Variational Equations and Tangent Space Evolution}
\label{subsec:variational}

Let the phase-space variables be denoted as \(\mathbf{X} = (z, p_z, x, p_x)\), with equations of motion given by Eqs.~\eqref{eq:eom1}-\eqref{eq:eom4}. Along a reference orbit \(\mathbf{X}(t)\), the evolution of a perturbation vector \(\mathbf{u}\) in the tangent space satisfies the variational equation:
\begin{equation}
\dot{\mathbf{u}} = \mathbf{J}(\mathbf{X}(t)) \cdot \mathbf{u},
\end{equation}
where \(\mathbf{J}\) is the Jacobian matrix of the equations of motion. We simultaneously follow a set of linearly independent tangent vectors \(\{\mathbf{u}_1, \dots, \mathbf{u}_4\}\), initialized as an orthonormal basis, whose collective evolution is described by a matrix \(\mathbf{U}(t)\) satisfying:
\begin{equation}
\dot{\mathbf{U}} = \mathbf{J}(\mathbf{X}(t)) \cdot \mathbf{U}.
\end{equation}

\subsection{Gram--Schmidt Orthogonalization and Exponent Extraction}
\label{subsec:gram_schmidt}

To avoid numerical ill-conditioning (all vectors aligning with the direction of maximal stretching), we periodically (every fixed integration interval \(\tau\)) perform Gram--Schmidt orthogonalization on \(\mathbf{U}\), obtaining an orthogonal set \(\{\mathbf{v}_1, \dots, \mathbf{v}_4\}\) and the corresponding norms \(\|\mathbf{w}_i\|\) (the lengths before normalization during orthogonalization), see more details in~\cite{1996Numerical}. After the \(k\)-th orthogonalization, the cumulative estimate for the \(i\)-th Lyapunov exponent is:
\begin{equation}
\lambda_{L_i}^{(K)} = \lim_{K\rightarrow{\infty}}\frac{1}{K\tau} \sum_{k=1}^{K} \ln \|\mathbf{w}_i^{(k)}\|.
\end{equation}
For sufficiently large \(K\), \(\lambda_{L_i}^{(K)}\) converges to the \(i\)-th Lyapunov exponent of the system, and the largest exponent \(\lambda_{L_1}\) is the MLE.

%\paragraph{Up to paragraphs.} We find that having more levels usually reduces the clarity of the article. Also, we strongly discourage the use of non-numbered sections (e.g.~\texttt{\textbackslash subsubsection*}).  Please also consider the use of ``\texttt{\textbackslash texorpdfstring\{\}\{\}}'' to avoid warnings from the \texttt{hyperref} package when you have math in the section titles.

\acknowledgments

This work is supported by the National Key Research and Development Program of China under Contract No. 2022YFA1604900, by the National Natural Science Foundation of China (NSFC) under Grants No. 12005033, No. 12564047, No. 12505151, No. 12435009, and No. 12275104. Si-wen Li is also supported by the Fundamental Research Funds for the Central Universities under Grant No. 3132026192. 

%\paragraph{Note added.} This is also a good position for notes added after the paper has been written.

\bibliographystyle{JHEP}
\bibliography{ref}

\end{document}